\documentclass[sn-mathphys-num]{sn-jnl}

\usepackage{graphicx}%
\usepackage{multirow}%
\usepackage{amsmath,amssymb,amsfonts}%
\usepackage{amsthm}%
\usepackage{mathrsfs}%
\usepackage[title]{appendix}%
\usepackage{xcolor}%
\usepackage{textcomp}%
\usepackage{manyfoot}%
\usepackage{booktabs}%
\usepackage{algorithm}%
\usepackage{algorithmicx}%
\usepackage{algpseudocode}%
\usepackage{listings}%

\theoremstyle{thmstyleone}%
\theoremstyle{thmstyletwo}%

\theoremstyle{thmstylethree}%

\begin{document}

\title[Article Title]{Non-additive Stochastic Model for Supercooled Liquids: New Perspectives for Glass Science}


\author*[1]{\fnm{Antonio Cesar} \sur{do Prado Rosa Jr.}}\email{antoniocprj@ufob.edu.br}

\author[1]{\fnm{Elias} \sur{Brito}}\email{eliasbaj@ufob.edu.br}
\equalcont{These authors contributed equally to this work.}

\author[1]{\fnm{Wanisson} \sur{Santana}}\email{wanisson.santana@ufob.edu.br}
\equalcont{These authors contributed equally to this work.}

\author[1]{\fnm{Clebson} \sur{Cruz}}\email{clebson.cruz@ufob.edu.br}
\equalcont{These authors contributed equally to this work.}

\affil*[1]{\orgdiv{Centro de Ci\^{e}ncias Exatas e das Tecnologias}, \orgname{Universidade Federal do Oeste da Bahia}, \orgaddress{\street{Rua Bertioga, 892}, \city{Barreiras}, \postcode{47810-059}, \state{Bahia}, \country{Brasil}}}


\abstract{We present a review of the Non-additive Stochastic Model for supercooled liquids  (NSM), an efficient approach for diffusive processes that provides a suitable interpretation for the non-Arrhenius dynamics in these materials. Based on a class of non-homogeneous continuity equations, the NSM provides functions able to model the thermal behavior of the viscosity and diffusivity in fragile liquids. The model defines a rigorous physical criterion that distinguishes super-Arrhenius from sub-Arrhenius processes, establishes a robust scale of fragility, describes fragile-to-strong curves in Angell's plot, and provides an accurate fitting equation for experimental viscosity data as effective as others viscosity classic models.}


\maketitle

\section{Introduction}\label{sec1}

The field of glass science is a rapidly developing field due to the wide range of technological applications for glass-forming liquids. However, the mechanisms behind the formation of amorphous solids are still not fully understood \cite{Berthier2011,Mauro2014.1,Mauro2014.2}. The glass-forming process occurs when a liquid undergoes a rapid cooling process reaching the supercooling metastable state that persists until the substance achieves its glass transition temperature. The glassy state shares a similar molecular arrangement to the supercooled liquid at the point of transition, with the fluid becoming highly viscous as it cools\cite{Berthier2011,Zheng2016}. The viscosity reflects a macroscopic dissipative effect resulting from local interactions among moving molecules and their surroundings. It provides significant information regarding the structural properties of the system that forms glass \cite{Rosa2020}. 

Thermally activated quantities, such as viscosity and diffusivity, characterize the diffusive processes in supercooled liquids. For some glass-forming liquids, activation energy is temperature-dependent, indicating non-Arrhenius-type processes. In this context, the glass-forming system is called \textit{strong liquid} if it follows the standard Arrhenius behavior \cite{Truhlar2001}, i.e., the activation energy is temperature independent, and viscosity and diffusivity are exponential functions of the inverse of the temperature. \textit{Fragile liquids} encompass the super-Arrhenius processes, for which viscosity and diffusivity curves correspond to non-exponential functions of the temperature, and the activation energy increases with the reciprocal temperature \cite{Truhlar2001,Rosa2016, Zheng2016,PhysRevE.100.022139}. In this context, we establish a model for the study of reaction-diffusion processes in glass-forming systems, named after the authors of the \textit{Non-additive Stochastic Model} (NSM), able to provide a suitable interpretation of the non-Arrhenius behavior in supercooled liquids \cite{PhysRevE.100.022139,Rosa2020,junior2024proofofconcept}.

The NSM is characterized by a class of non-homogeneous continuity equations (NCE) corresponding to the non-linear Fokker-Planck equation, whose generalized drag and diffusion coefficients are functional forms of the concentration function, solution of the NCE \cite{PhysRevE.100.022139}. The concentration corresponds to a class of rapidly decreasing functions whose stationary form maximizes nonadditive entropies \cite{Schwmmle2009,dosSantosMendes2017}, generalized drag coefficient involves information about dissipative or exchange effects, and the diffusion coefficient describes anomalous diffusion processes. The proposed approach furnishes a class of non-exponential functions for the description of the thermal behavior of diffusivity, viscosity, and the activation energy in fragile liquids, characterized by a threshold temperature associated with viscosity divergence in super-Arrhenius processes \cite{PhysRevE.100.022139}.  

In a later study \cite{Rosa2020}, we analyzed the NSM viscosity equation and replicated Angell's plot \cite{Angell1988,Angell2002}, which is a standard method used to classify fragile liquids. We observed a direct relationship between the fragility index and the activation energy of glass transition temperature. This finding provides a reliable method to evaluate the level of fragility in glass-forming systems \cite{Rosa2020}. Additionally, we used the two-state model \cite{Tanaka2000,Shi2018} to simulate the fragile-to-strong transition curve in Angell's plot, varying the threshold temperature of viscosity divergence to explore this transition.

In our latest publication, we demonstrate a proof-of-concept of the Non-Stationary Model (NSM) in supercooled liquids \cite{junior2024proofofconcept}. This is achieved by analyzing the temperature-dependent viscosity data of twenty-five different types of glass-forming materials. We use the viscosity equation to model the thermal response of the activation energy and ascertain the values of the glass transition temperature and the fragility index for each substance. Finally, we demonstrate the accuracy of the NSM compared to the Vogel-Fulcher-Tammann (VFT),  Mauro-Yue-Ellison-Gupta-Allan (MYEGA), and Avramov-Milchev (AM) equations \cite{Mauro2009, Zhu2018}, other models for the study of temperature-dependent viscosity in fragile liquids.

In this paper, we present the state-of-art of the Nonadditive Stochastic Model for studying the dynamic properties in supercooled liquids, discussing its theoretical scope and the main results obtained to date, based on references \cite{PhysRevE.100.022139,Rosa2020,junior2024proofofconcept}. Furthermore, this review aims to illustrate how this methodology emerges in the literature as a reliable tool for characterizing the non-Arrhenius behavior in glass-forming systems \cite{CarvalhoSilva2019,CarvalhoSilva2020,Kohout2021,Emran2022,Roy2022,Bondarchuk2023}.

\section{Non-additive Stochastic Model (NSM)}\label{sec2}

\begin{figure}[H]
	\begin{center}
		\includegraphics[width=8cm]{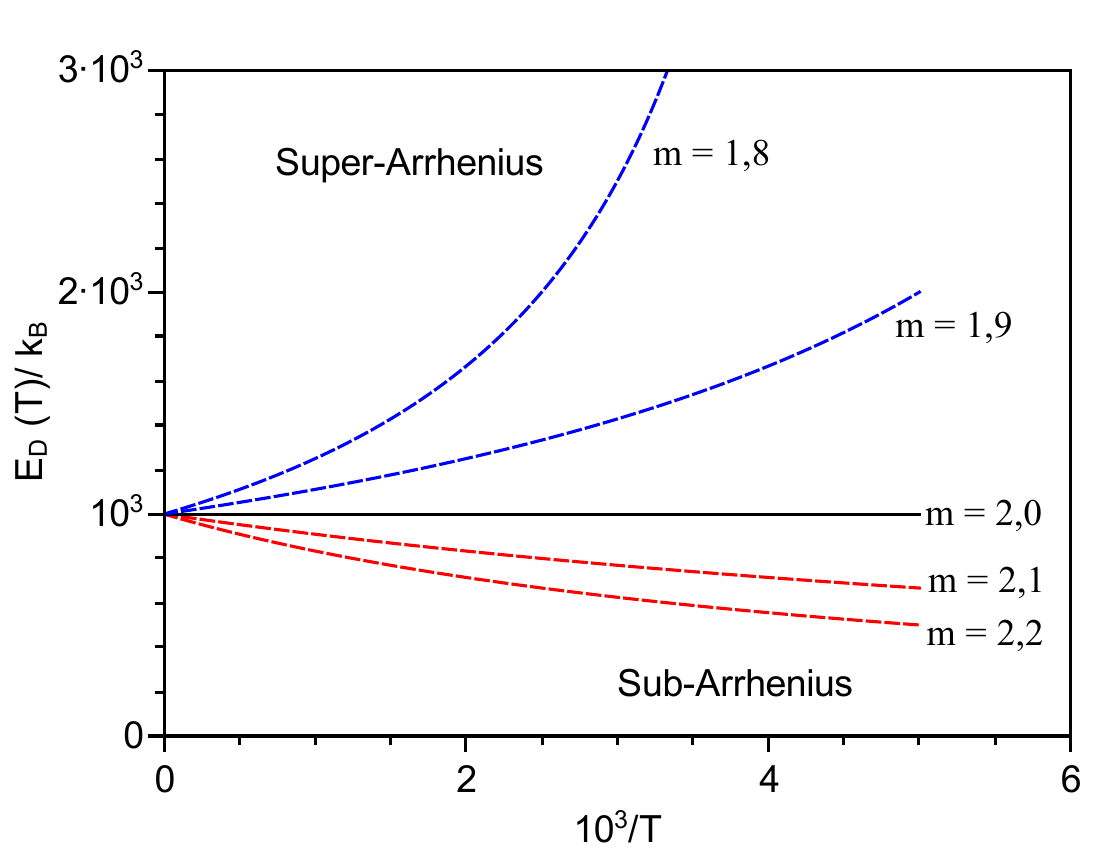}
		\caption{Activation energy in terms of the reciprocal temperature. When the curves show activation energy as a decreasing function of the reciprocal temperature, it illustrates a class of sub-Arrhenius processes. On the other hand, if the curves show an increasing activation energy as a function of the reciprocal temperature, it describes a class of super-Arrhenius processes. The curve where $m=2$ characterizes the Arrhenius activation energy, which is described by a temperature-independent behavior. These plots were produced using the temperature scale factor $E / kB = 1000 K$ \cite{PhysRevE.100.022139}.}
		\label{figure1}
	\end{center}
\end{figure}

In the NSM, the concentration $\rho(r,t)$ of a glass-forming liquid is the solution of a non-homogeneous continuity equation, which refers to dissipative reaction-diffusion processes \cite{PhysRevE.100.022139}. In this case, the diffusion flux is a generalized version of Fick's first law, where the generalized diffusion coefficient is proportional to $\rho(r,t)$. The volumetric density per unit of time associated with dissipative processes corresponds to the divergence of a dissipation field proportional to $\rho^m \vec \nabla \phi$, where $\phi$ is a generalized chemical potential and $m$ is a characteristic exponent of NSM. Under such conditions, the NCE corresponds to a class of nonlinear Fokker-Planck equation, given by,
\begin{equation}
	\frac{\partial \rho(r,t)}{\partial t}= \kappa_{m}^{-1}\vec{\nabla} \cdot\left[ \left( \vec{\nabla} \phi\right) \rho^{m}\right] + \frac{\Gamma}{2}{\nabla}^2 \left[ \rho^2\right]
	\label{Eq1}
\end{equation}
where $ \kappa_{m}$ and $\Gamma$ are positive proportionality constants. The stationary solution of the Eq. \eqref{Eq1} is a non-exponential density function that maximizes nonadditive entropic forms, such as the Tsallis entropy\cite{Schwmmle2009,dosSantosMendes2017}, and this condition we can describes the temperature dependent diffusivity as,
\begin{equation}
	D(T) = D_0 \left[ 1-(2-m)\frac{E}{k_B T} \right]^{\frac{1}{2-m}}
	\label{Eq2}
\end{equation}
where $D_0$ is the diffusivity in the high-temperature limit, $k_B$ is the Boltzmann constant, and $E$ is a potential barrier associated with $\phi$. Experimental diffusivity data are satisfactorily fitting for a diffusion coefficient similar to Eq.\eqref{Eq2} in \cite{Rosa2016}. Applying the definition of Arrhenius law to Eq. \eqref{Eq2}, the diffusivity activation energy is given by: 
\begin{equation}
	E_{D}(T) = \frac{E}{1-(2-m)\frac{E}{k_B T}}
	\label{Eq3}
\end{equation}
Figure \ref{figure1} shows $E_{D}(T)$ behavior for different $m$ values. The curves $m < 2$ (blue dashed lines) correspond to the class of super-Arrhenius processes, for which the activation energy increases as the supercooled liquid approaches the glass transition. In this case, the activation energy diverges, and the diffusivity disappears for the threshold temperature $T_t = (2-m)E/k_B$. For $m > 2$ in Eq.3, the activation energy curves (red dashed lines) decreases as the temperature drops, defining sub-Arrhenius type processes.. Arrhenius law is recovered for condition $m =2$ (black solid line) and $E_D = E$.
Additionally, a temperature-dependent viscosity equation for supercooled liquids emerges from the NSM formalism. In Eq. \eqref{Eq1}, the generalized fluid mobility associated with the drift velocity to the dissipation field corresponds to $\kappa_m^{-1} \rho^{m-1}$, and is proportional with the inverse of the viscosity \cite{PhysRevE.100.022139,Rosa2020}. For the steady state of the Eq. \eqref{Eq1}, this approach provides a non-exponential function for the thermal behavior of viscosity $\eta(T)$,  given by,
\begin{equation}
	\eta(T) = \eta_\infty \left[ 1-(2-m)\frac{E}{k_B T} \right]^{-\gamma}
	\label{Eq4}
\end{equation}
where $\eta_\infty$ the high-temperature viscosity limit, and $\gamma = (m-1)/(2-m)$ \cite{PhysRevE.100.022139,Rosa2020}. Eq.\eqref{Eq4} is an increasing function of reciprocal temperature for super Arrhenius processes ($m<2$), assuming large values as it approaches the glass transition and diverging for the temperatures close to $T_t$. However, by convention, we have that the glass transition temperature ($T_g$) occurs for $\eta = 10^{12}$ Pa.s, implying that $T_g > T_t$ \cite{PhysRevE.100.022139,Rosa2020}. In Eq.\eqref{Eq4}, the standard Arrhenius behavior is recovered by $m = 2$, and sub-Arrhenius processes corresponds to $m>2$.

\begin{figure}[H]
	\begin{center}
		\includegraphics[width=8cm]{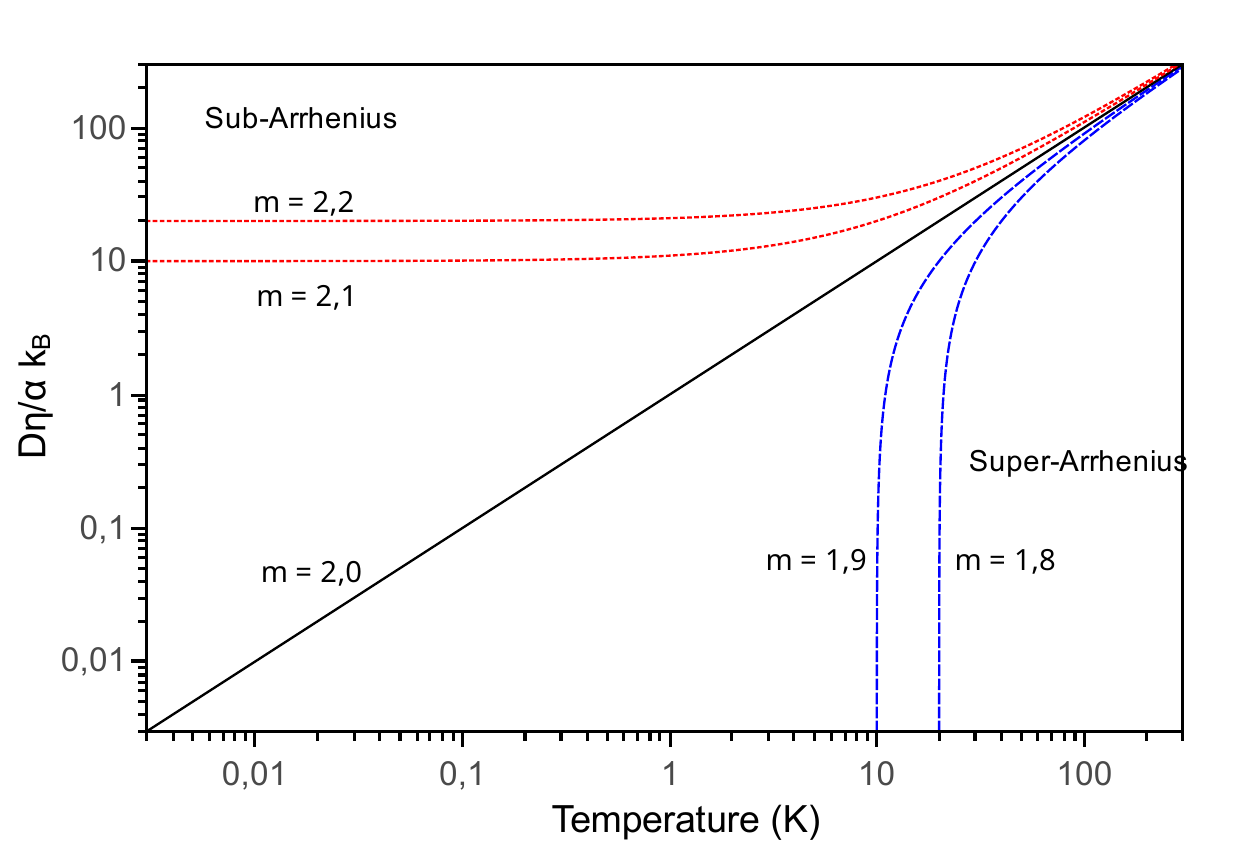}
		\caption{The generalized Stokes-Einstein relation in function of the temperature (logarithmic scale), Eq.\eqref{Eq5}. Curves for $m < 2$ (dashed blue lines) corresponds to super-Arrhenius processes, and rapidly goes to zero as $T \rightarrow T_t$. The sub-Arrhenius processes corresponds to $m > 2 $ (dotted red lines), and a temperature independent behavior occurs from condition $E \gg k_B T$. The usual Stokes-Einstein relation (straight line), associated with the standard Arrhenius process, occurs for $m = 2$ or the condition $E \ll k_B T$ \cite{PhysRevE.100.022139}.}
		\label{fig2}
	\end{center}
\end{figure}

Based on the fluctuation-dissipation theorem, the NSM provides a generalized version of the Stokes-Einstein relation that considers non-Arrhenius processes \cite{PhysRevE.100.022139}. Thus, from Eq.\eqref{Eq1} and Eq.\eqref{Eq4}, we obtain,
\begin{equation}
	D \eta = \alpha k_B T \left[ 1-(2-m)\frac{E}{k_B T} \right]
	\label{Eq5}
\end{equation}
where $\alpha$ is a positive definite constant. As illustrated in Figure \ref{fig2}, for super-Arrhenius processes ($m<2$), Eq.\eqref{Eq5} decreases rapidly as the supercooled liquid approaches the glass transition, and the generalized Stokes-Einstein relationship nullifies it for $ T=T_t$. On the other hand, the condition $E \gg k_B T$ in sub-Arrhenius processes ($m>2$) implies a temperature-independent behavior for Eq.\eqref{Eq5}. Finally, the usual Stokes-Einstein relation is recovered both for $m=2$ and for the condition $E \ll k_B T$ whatever the value of $m$. In this way, the NSM provides a robust physical criterion for distinguishing different types of non-Arrhenius processes.

\section{Fragility and Fragile-to-Strong transition in the NSM}\label{sec3}

Fragility can be defined as a macroscopic parameter that characterizes the degree of short-range order present in the molecular arrangement of a supercooled liquid during the glass transition. The higher the fragility index, the lower the presence of a well-defined short-range order \cite{Zhu2018}. Angell's plot is a widely used technique for categorizing the fragility of glass-forming liquids \cite{Angell2002,Zheng2016,Rosa2020}. It examines the dependence between the logarithm of viscosity and the $T_g/T$ ratio, where the standard Arrhenius behavior corresponds to a straight line, and fragile liquid curves are characterized by a deviation from linearity\cite{Angell2002,Zheng2016,Rosa2020}. Thus, the degree of fragility of a supercooled liquid corresponds to the slope of Angell's plot for $T = T_g$, named the fragility index ($M_\eta$). Applying Angell's definitions to Eq.\eqref{Eq4}, we obtain:
\begin{equation}
	M_\eta = \gamma \frac{T_t }{T_g-T_t }
	\label{Eq6}
\end{equation}

\begin{figure}[H]
	\begin{center}
		\includegraphics[width=8cm]{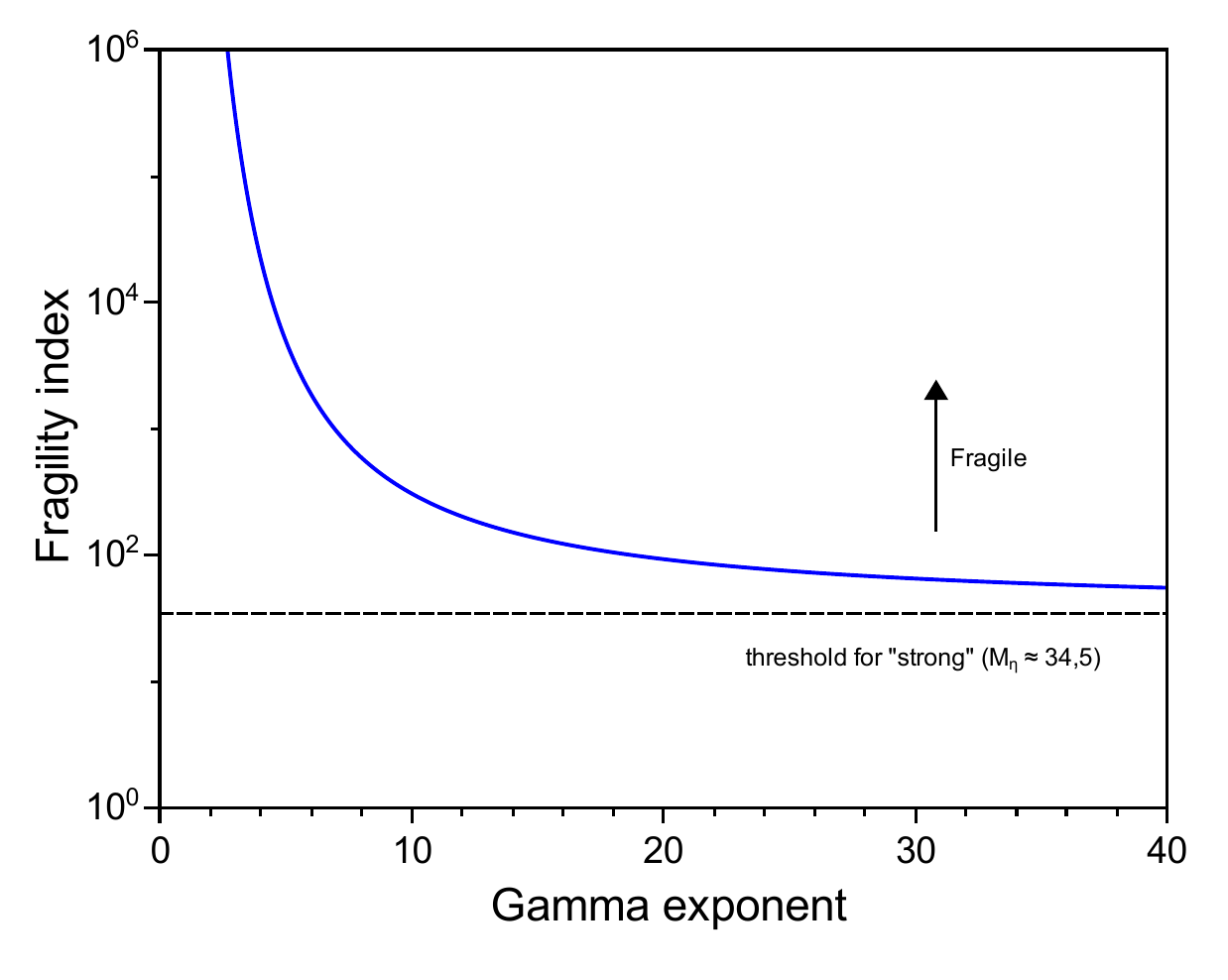}
		\caption{Eq.\eqref{Eq7} for $B=15$. The solid blue line shows the dependence of the fragility index $M_\eta$ with the $\gamma$ exponent. The threshold between fragile and strong regimes (horizontal line) corresponds to the value $M_\eta \approx 34,5$.}
		\label{fig3}
	\end{center}
\end{figure}

According to Eq.\eqref{Eq6}, the values difference between $T_t$ and $T_g$ conditions the fragility index of the glass-forming liquid, so that the smaller the difference, the greater the degree of fragility of the liquid. Alternatively, the fragility index corresponds to the ratio between the viscosity activation energy (given by $(1-m) E_D$, see Eq.\eqref{Eq3}) and the thermal energy $k_B T$ for $T=T_g$ \cite{Rosa2020}. In this context, the  NSM provides a unique relation between $\gamma$ exponent and the fragility index, written as,
\begin{equation}
	M_\eta = \gamma \left( 10^{B/\gamma}-1 \right)
	\label{Eq7}
\end{equation}
where $B = 12 - log_{10}\eta_\infty$ \cite{Rosa2020}.   From Eq.\eqref{Eq7}, the fragility index is a decreasing function of the $\gamma$ exponent, and the threshold between fragile and strong behavior occurs for the asymptotic limit $\gamma \rightarrow \infty$ (equivalent to $m \rightarrow 2$), which implies $M_\eta \rightarrow B ln (10)$\footnote{In \cite{Rosa2020}, we neglect the constant $ln(10)$.} \cite{junior2024proofofconcept}.  Figure \ref{fig3} shows $M_\eta$ as a function of $\gamma$ for $B=15$, considering $log_{10} \eta_\infty = -3$ due to experimental evidences of the universal behavior for $\eta_\infty$\cite{Rosa2020,junior2024proofofconcept}.  Therefore, it is seen that through the $\gamma$ exponent , the NSM provides a reliable parameter to classify the degree of fragility of a glass-forming liquid.

\begin{figure}[H]
	\begin{center}
		\includegraphics[width=8cm]{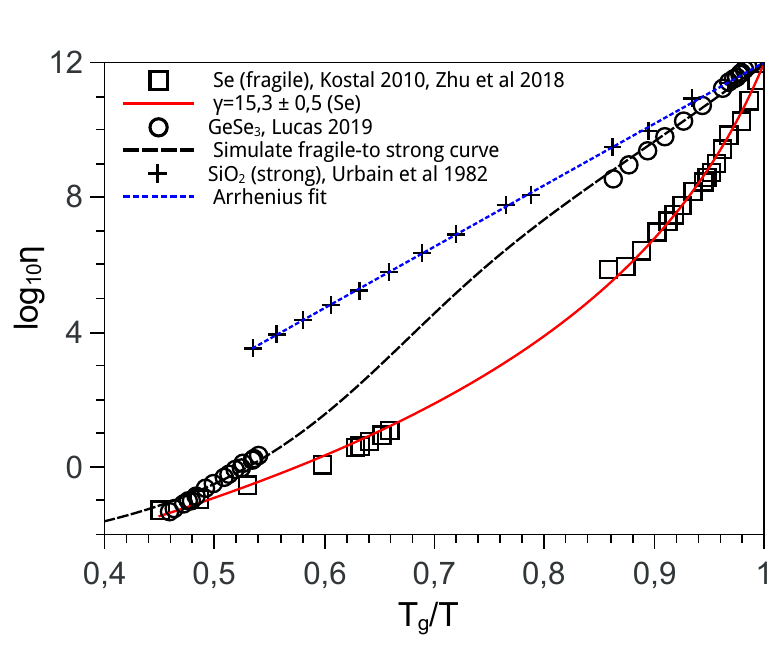}
		\caption{Angell plot for experimental results and the corresponding fits produced by NSM. The dashed red line depicts the fit of the fragile liquid shown by the selenium melt (open squares). The dashed red line represents the fit of the fragile liquid behavior demonstrated by the selenium melt (red open squares). The dashed blue line depicts the linear regression of the classic Arrhenius behavior found for liquid silica, SiO$_{2}$ ( shown by black crosses).  The dashed black line represents the estimated transition from fragile to strong of the GeSe$_{3}$ glass-forming system in comparison to the viscosity values given in the literature (shown by black open circles). \cite{Rosa2020}}
		\label{fig4}
	\end{center}
\end{figure}

The classification as fragile or strong liquid does not apply for some glass-forming systems, given that its viscosity curves undergo a transition in Angell's plot, indicating a switch-over of the super-Arrhenius regime for standard Arrhenius, which does characterize the fragile-to-strong transition phenomenon \cite{Shi2018,Lucas2019,PhysRevE.100.022139,Rosa2020}. In NSM, changes in the dynamic properties in the supercooled liquid that undergoes the fragile-to-strong transition correspond to a variation of the potential barrier $E$, consequently threshold temperature $T_t$, during the glass formation process. This condition implies that the viscosity activation energy converges to a constant value as it approaches the glass transition temperature.  We implement the two-state model to characterize the potential barrier $E$ variation during the glass transition process and simulate the fragile-to-strong curve in Angell's plot for the glass-forming system $GeSe_3$, as shown in Figure \ref{fig4}.

The standard Arrhenius behavior (strong liquid) is characterized by the linear fit (blue dashed line) about experimental viscosity data (black crosses) of silica liquid ($SiO_2$). The red solid line is the fit of Eq. \eqref{Eq4} and describes the selenium melt ($\gamma = 15,3$) as a fragile liquid.  From experimental viscosity data (black open circles) of the $GeSe_3$ glass-forming system, we simulate the fragile-to-strong curve (black dashed line), considering for the two-state model $\gamma = 4,05$ for the fragile liquid regime (state I) and $\gamma = 37,69$ for the strong liquid regime (state II).

\section{Proof-of-concept for the NSM in  supercooled liquids}\label{sec4}

\begin{figure}[H]
	\begin{center}
		\includegraphics[width=8cm]{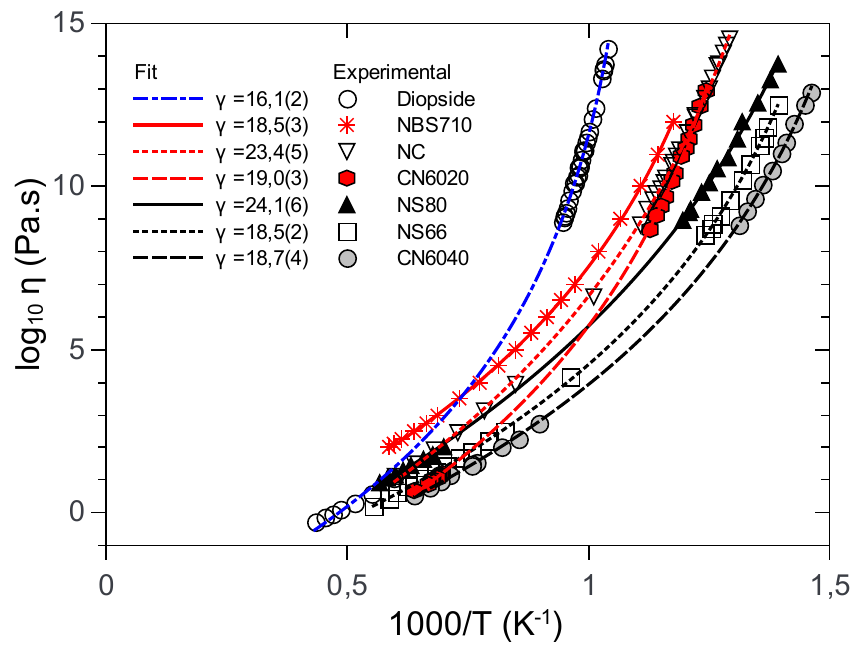}
		\caption{Thermal behavior of viscosity for silicate glasses. The different symbols describe the experimental data \cite{Urbain1982,Sipp2001, Neuville2006,Jaccani2017}, and the curves (continuous and dashed) depict the nonlinear fit of the logarithm in Eq.\eqref{Eq4} \cite{junior2024proofofconcept}.}
		\label{fig5}
	\end{center}
\end{figure}

In recent work, we apply the nonlinear regression fitting of the Eq.\eqref{Eq4} for dependent-temperature viscosity data from twenty-five types of glass-forming systems, corresponding to silicates, borosilicates, aluminosilicates, titania silicates, and chalcogenide glasses \cite{junior2024proofofconcept}. The fit of the NSM viscosity equation provides $\chi^2$  test values less than $0,1$ and Pearson coefficient values in order of $0,999$ for all substances, demonstrating that Eq.\eqref{Eq4} is an efficient model to characterize the thermal behavior of viscosity in glass-forming liquids. In a logarithmic scale, The graph depicted in Figure \ref{fig5} illustrates the viscosity of silicate glasses as a function of reciprocal temperature. The continuous and dashed lines on the graph represent the fitted curves of the Eq. \eqref{Eq4} experimental viscosity data, which is presented by various symbols. For more information on other substances, refer to Figure 5 in ref. \cite{junior2024proofofconcept}.

From the fit parameters, see Table I in \cite{junior2024proofofconcept}, we simulate the temperature-dependent activation energy of viscosity, begin an increasing function of the reciprocal temperature for all substances (see Figure 2 and Figure 6 in \cite{junior2024proofofconcept}), characterizing the super-Arrhenius behavior. For each glass-forming substance, we calculated the glass transition temperature and fragility index values (see Table I in \cite{junior2024proofofconcept}).  The relation between $M_\eta$ and $\gamma$ exponent values (see Figure 3 in \cite{junior2024proofofconcept}) corroborate with Eq.\eqref{Eq7}, where the dispersion of data related to the theoretical plot, Eq. \eqref{Eq7}, is a consequence of the experimental high-temperature viscosity limit obtained for each glass-forming liquid.

\begin{figure}[H]
	\begin{center}
		\includegraphics[width=8cm]{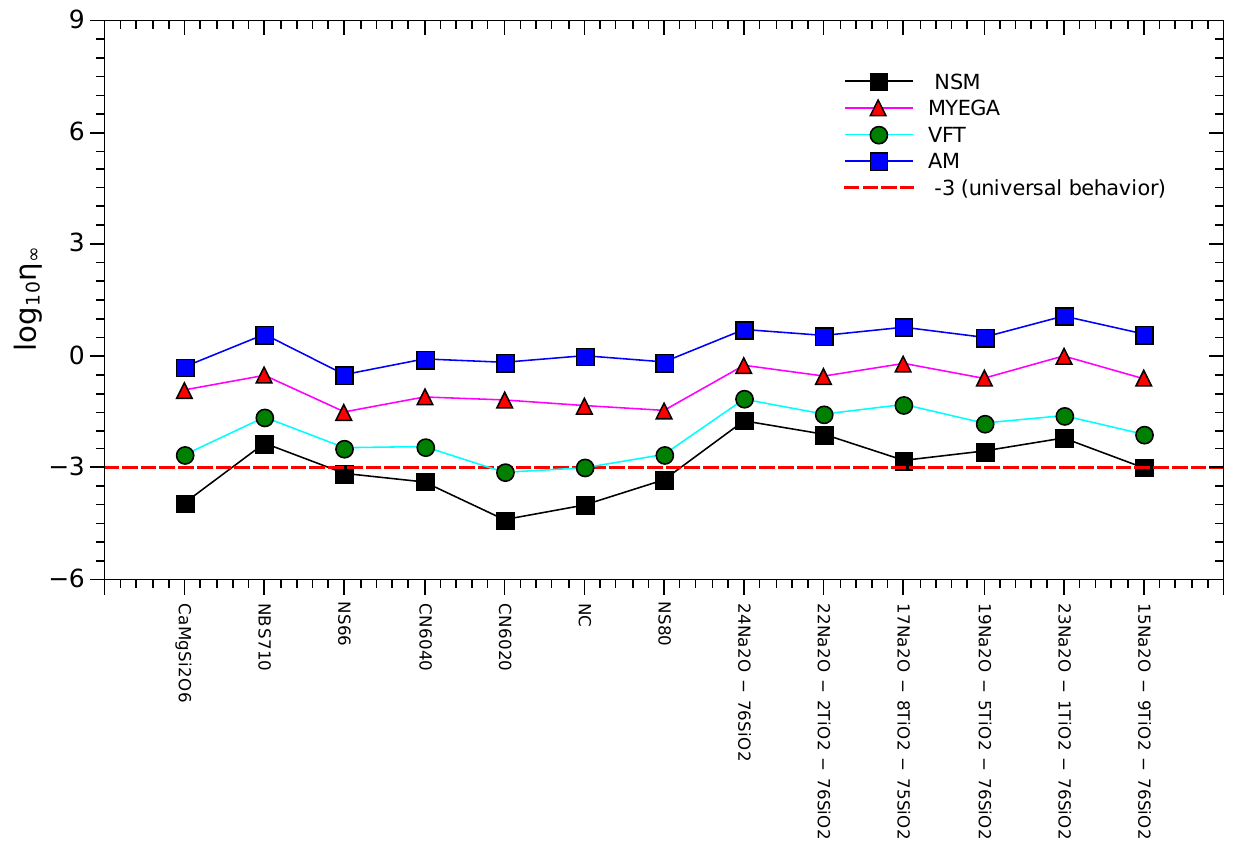}
		\caption{Values of $log_{10} \eta_\infty$ for silicates and titania-silicates glasses obtained from the AM (blue squares), MYEGA (purple triangles), and VFT (green circles) models. The black squares corresponds to the NSM values. The dashed red line is the universal behavior $log_{10} \eta_\infty = -3$ \cite{junior2024proofofconcept}.}
		\label{fig6}
	\end{center}
\end{figure}

We compare the NSM results with the Vogel-Fulcher-Tammann (VFT),  Avramov-Milchev (AM), and Mauro-Yue-Ellison-Gupta-Allan (MYEGA) models, which provide temperature-dependent viscosity equations serviceable for modeling fragile liquids \cite{Sipp2001,Zhu2018}. These models generate $\chi^2$ and $R^2$ values in the same order of magnitude as the results obtained from Eq.\eqref{Eq4} (see Tables I and II in \cite{junior2024proofofconcept}), emphasizing the NSM accuracy for characterizing the thermal behavior of viscosity in glass-forming systems. Finally, we saw that Eq.\eqref{Eq4} provides values close to the $10^{-3}$ Pa.s for the high-temperature viscosity limit in silicate and titania-silicate glasses, as the VFT, AM, and MYEGA models overestimate $\eta_\infty$ for same substances, according to shows Figure \ref{fig6}.  

\section{Conclusion}\label{sec13}

	In summary, the NSM establishes a class of non-homogeneous continuity equations to describe reaction-diffusion processes in glass-forming liquids, providing non-exponential functions of the diffusivity and viscosity and a temperature-dependent activation energy, which interpret the non-Arrhenius behavior in supercooled liquids, during the glass transition process. The $m$ exponent is an efficient parameter to differentiate super-Arrhenius processes ($m<2$), typical in fragile liquids and associated with classical transport phenomena, from sub-Arrhenius processes ($m>2$), associated with non-local quantum effects \cite{PhysRevE.100.022139}, and recovered the standard Arrhenius law for $m=2$ condition. Furthermore, the generalized Stokes-Einstein relation that emerges from the model establishes a rigorous physical criterion that distinguishes the different non-Arrhenius-type processes. 
	
	Fragile liquids present a threshold temperature $T_t$ for which the viscosity and activation energy diverge, the diffusivity is null, and the $T_t/T_g$ ratio is a robust indicator of the fragility in supercooled liquids. The fragility index $M_\eta$ is associated with the viscosity activation energy for $T=T_g$, and the $\gamma$ exponent of the NSM viscosity equation is a reliable parameter to define the degree of fragility in glass-forming systems, given Eq.\eqref{Eq7}. The fragile-to-strong transition corresponds to structural changes in the molecular arrangement of the supercooled liquid during the glass formation process, corresponding to the variation of the threshold temperature $T_t$. From this condition, we show that the NSM describes the fragile-to-strong transition curve for the glass-forming system $GeSe_3$.
	
	After analyzing experimental data from twenty five types of glass-forming systems and comparing it to other models, including the NSM, VFT, AM, and MYEGA, we have found that Eq. \eqref{Eq4} accurately models the thermal behavior of viscosity in fragile liquids. We also verified that for silicate and titania-silica glasses, the NSM provides values for $\eta_\infty$ close to the universal behavior $10^{-3}$ Pa.s, while the other models overestimated the same parameter.
	
	The results presented in this review unequivocally demonstrate that the Non-Additive Stochastic Model is an efficient approach for studying dynamic properties in supercooled liquids and characterizing non-Arrhenius behavior in glass-forming systems. Furthermore, the relationship between the stationary solution of the NCE and non-additive entropic forms defines a fruitful path for studying the thermodynamic properties of these systems, which is the topic for future work.

\backmatter

\bmhead{Supplementary information}

\bmhead{Acknowledgements}
The authors would like to thank the Bahia State Research Support Foundation (FAPESB) for its financial support.



\end{document}